\documentclass[nocompress]{spie}  %>>> to avoid compression of citations

\usepackage{amsmath,amsfonts,amssymb}
\usepackage{graphicx}
\usepackage[colorlinks=true, allcolors=blue]{hyperref}

\usepackage{subcaption}
\usepackage{multirow}
\usepackage{bm}
\usepackage[labelfont=bf, skip=8pt]{caption}
\usepackage[indLines=false]{algpseudocodex}

\usepackage{fancyhdr}
\usepackage{booktabs}

\def\PAnumber{AFRL-2024-4421}
\fancypagestyle{empty}{%
  \fancyhf{} 
  \fancyfoot[C]{Approved for public release; distribution is unlimited.  Public Affairs release approval \# \PAnumber .}
  
}

\title{Data driven synthetic wavefront generation for boundary layer data}

\author[a]{Jeffrey W. Utley}
\author[a]{Gregery T. Buzzard}
\author[b]{Charles A. Bouman}
\author[c]{Matthew R. Kemnetz}
\affil[a]{Department of Mathematics, Purdue University, West Lafayette, Indiana 47907, USA}
\affil[b]{Departments of Electrical and Computer Engineering, and Biomedical Engineering, Purdue University,  West Lafayette, Indiana 47907, USA}
\affil[c]{Directed Energy Directorate, Air Force Research Laboratory, Kirtland Air Force Base, New
Mexico 87117, USA}

\begin{document} 
\maketitle

\begin{abstract}
    Disturbances such as atmospheric turbulence and aero-optic effects lead to wavefront aberrations, which degrade performance in imaging and laser propagation applications. Adaptive optics (AO) provide a method to mitigate these effects by pre-compensating the wavefront before propagation. However, development and testing of AO systems requires wavefront aberration data, which is difficult and expensive to obtain. Simulation methods can be used to generate such data less expensively. For atmospheric turbulence, the Kolmogorov-Taylor model provides a well-defined power spectrum that can be combined with the well-known angular spectrum method to generate synthetic phase screens. However, as aero-optics cannot be similarly generalized, this process cannot be applied to aero-optically relevant phenomena. In this paper, we introduce ReVAR (Re-Whitened Vector Auto-Regression), a novel algorithm for data-driven aero-optic phase screen generation. ReVAR trains on an input time-series of spatially and temporally correlated wavefront images from experiment and then generates synthetic data that captures the statistics present in the experimental data. The first training step of ReVAR distills the input images to a set of prediction weights and residuals. A further step we call re-whitening uses a spatial principal component analysis (PCA) to whiten these residuals. ReVAR then uses a white noise generator and inverts the previous transformation to construct synthetic time-series of data. This algorithm is computationally efficient, able to generate arbitrarily long synthetic time-series, and produces high-quality results when tested on turbulent boundary layer (TBL) data measured from a wind tunnel experiment. Using measured data for training, the temporal power spectral density (TPSD) of data generated using ReVAR closely matches the TPSD of the experimental data.
\end{abstract}


\section{INTRODUCTION}
\label{s: introduction}
Atmospheric turbulence and aero-optic effects cause disturbances in propagating light waves. Such disturbances are characterized as wavefront aberrations, in which the shape of a planar wavefront is distorted. Wavefront aberrations degrade laser beam propagation from an aircraft, as well as inhibit imaging applications. Adaptive-optic (AO) systems mitigate these issues by employing a deformable mirror (DM) to pre-compensate the wavefront prior to propagation. Development of AO control algorithms typically require wavefront aberration data to test the effectiveness of DM correction. However, physical wavefront aberration data is expensive and difficult to obtain. Simulation methods could provide a much cheaper alternative for acquiring such data. In the case of atmospheric turbulence, the Kolmogorov-Taylor model accurately describes the prevailing physics.  More specifically, the Kolmogorov-Taylor model provides a well-defined power spectrum, which can be combined with the well-known angular spectrum method to generate relevant phase screens. However, aero-optic effects exhibit significantly more complicated statistical properties which the Kolmogorov-Taylor model is unable to capture. Therefore, we developed a novel data-driven method for generating aero-optics phase screens. 

Aero-optics is a field of study concerned with the interaction of light with aerodynamic flows. More specifically, a turbulent flow field causes density variations within a given flow field. As a result of these density fluctuations, the refractive index within the flow field varies as a function of both position and time. Light waves propagated throughout the flow field will then experience wavefront aberrations. These aberrations are often referred to as aero-optic aberrations and are quantified using optical path difference (\textit{OPD}) \cite{annurev:/content/journals/10.1146/annurev-fluid-010816-060315}. For monochromatic light (e.g., laser beams), \textit{OPD} is proportional to the phase errors induced by refractive index variations.

A significant issue posed by aero-optics arises in laser beam propagation from an aircraft. Notably, aerodynamic flow over an aircraft induces a turbulent boundary layer (TBL) near the beam director \cite{annurev:/content/journals/10.1146/annurev-fluid-010816-060315}. This TBL induces wavefront aberrations in the light propagated from this turret. These aberrations scale temporally with the air speed of the aircraft. As the aircraft moves faster, the bandwidth requirements for traditional adaptive-optic compensation are increased. As a result, research into novel AO control algorithms has flourished within the aero-optics community.

Physically relevant wavefront aberration data is necessary for the development of AO algorithms focused on mitigating the aero-optics problem. While wind tunnel \cite{Kemnetz-2016-SanDiego}, flight experiments \cite{jumperzenk-grodeyev}, and high fidelity computational fluid mechanics (CFD) simulations \cite{doi:10.1243/09544100JAERO385, WANG2013411} have previously captured these quantities, they are expensive and time-intensive for the amount of data they provide. Thus, synthetic data generation techniques provide a more efficient data acquisition method. In the case of atmospheric turbulence, the Kolmogorov-Taylor model gives us a well-defined power spectrum which angular spectra methods use to generate physically relevant synthetic phase screens \cite{lens.org/071-211-030-479-936}. However, application of this model relies on (spatial) homogeneity and isotropy, properties that atmospheric turbulence follows. Aero-optic effects exhibit complicated statistical properties, making the process neither homogeneous nor isotropic \cite{Vogel:14}. Because the Kolmogorov-Taylor model thus does not provide a well-defined power spectrum for aero-optic effects, the synthetic generation method described by Schmidt \cite{lens.org/071-211-030-479-936} does not apply. As a result, alternative synthetic generation methods are necessary for aero-optic effects.

In this paper, we introduce the ReVAR (Re-Whitened Vector AutoRegression) algorithm, a novel algorithm designed to generate synthetic wavefront data capturing spatial and temporal properties of measured data. This algorithm first analyzes a time-series of spatially and temporally correlated images. Using a white noise generator, ReVAR then creates synthetic data with the same spatial and temporal statistics as the input data. Importantly, ReVAR constructs synthetic data on arbitrarily long time-scales. This model includes the following novel techniques:
\begin{itemize}
    \item ReVAR uses a higher-order predictive linear model to capture temporal statistics.
    \item The model performs spatial whitening and vectorized linear time prediction independently to analyze both the spatial and temporal statistics of the input data.
    \item This algorithm introduces a re-whitening step which uses a second layer of spatial whitening to address the spatial effects of vectorized linear time prediction.
\end{itemize}
When applied to measured aero-optics data \cite{Kemnetz-2016-SanDiego}, ReVAR creates high-quality aero-optics simulations. Furthermore, the algorithm is computationally inexpensive and easy to run. As a result, ReVAR provides an efficient alternative to existing simulation algorithms for aero-optic effects.

\section{METHODOLOGY}
\label{s: Methodology}

ReVAR uses a white noise generator to create synthetic data on arbitrary time-scales which matches the spatial and temporal statistics of input data. The model first estimates statistics from the input data and uses these quantities to transform the data into white noise $N(\mathbf{0}, I)$. Crucially, the resulting transformation is invertible, allowing ReVAR to then convert a time-series of white noise into a time-series of synthetic data (with the same length). Because white noise generation is computationally inexpensive, ReVAR uses this approach to efficiently generate a long time-series of synthetic data.

In capturing the spatial and temporal statistics of the input data, ReVAR independently applies two distinct estimation tools. Specifically, the model uses Principal Component Analysis (PCA) to capture the spatial statistics and Vector AutoRegressive (VAR) modeling to extract temporal statistics. These methods are chosen because they each contribute important transformations in the conversion from input data to white noise: PCA for spatial whitening and VAR modeling for vectorized linear time prediction. Further, the inverses of these transformations also incorporate the input data's spatial and temporal statistics to previously un-correlated data. Thus, distinct application of PCA and VAR modeling allows ReVAR to both extract and integrate spatial and temporal statistics.

ReVAR is naturally split into two distinct steps: data analysis and synthesis. In performing data analysis, ReVAR estimates the input data's spatial and (short-range) temporal statistics. From these statistics, ReVAR then constructs the transformation which converts the input time-series to white noise. To perform synthesis, ReVAR first uses a white noise generator to create a time-series of white noise with some desired $N_s$ time-steps. ReVAR then applies the inverse transformation to the white noise into a time-series of synthetic data images which captures the desired spatial and (short-range) temporal correlation. After applying this transformation, ReVAR performs a post-processing step to incorporate the long-range temporal correlation of the input data. The final result of these two steps is a time-series of synthetic data images (of length $N_s$) which matches the spatial and temporal statistics of the input data.

\section{RESULTS}
\label{s: Results}

We tested the ReVAR algorithm on TBL data and used temporal power spectral density (TPSD) calculations to assess the model's performance. The data sets F06 and F12 resulting from the experiment described in \cite{Kemnetz-2016-SanDiego} provide two distinct time-series of images containing measured aero-optics data. We used each data set as input to ReVAR and thus generated two time-series of synthetic aero-optics data. In order to determine how accurately the algorithm captures the measured data's statistics, we compared the TPSD of the input data to the TPSD of the synthetic data. Our results show high quality TPSD fit for both data sets.

We use experimental aero-optics data to train and test ReVAR. The experimental data comes from the work of Kemnetz and Gordeyev \cite{Kemnetz-2016-SanDiego}, who induced a TBL on laser beams propagated through a wind tunnel and measured the subsequent wavefront aberrations using a Shack-Hartmann Wavefront Sensor (SHWFS). The result of this TBL data is a time-series of spatially and temporally correlated images whose pixels contain \textit{OPD} values. Importantly, tip, tilt, and piston (TTP) were removed from the measurements to reduce the influence of aero-mechanical aberrations in the experimental data \cite{Kemnetz-2016-SanDiego}.

\begin{figure}[t]
    \centering
    \begin{subfigure}{0.45\textwidth}
        \centering
        \includegraphics[width=\textwidth]{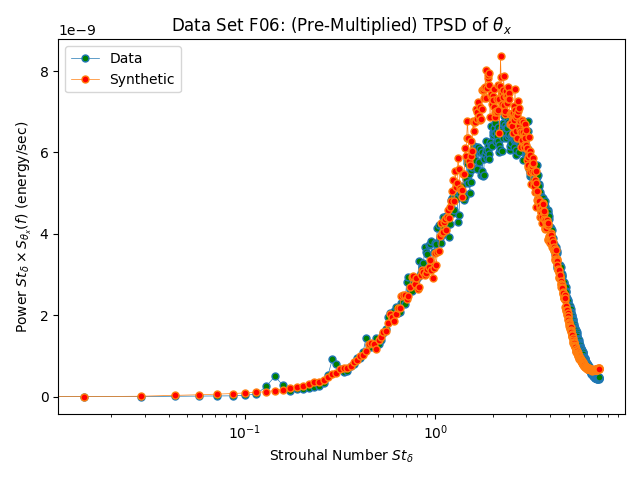}
        \caption{Data Set F06: TPSD of $\theta_x$}
        \label{fig: F06 TPSD of Deflection Angle}
    \end{subfigure}
    \begin{subfigure}{0.45\textwidth}
        \centering
        \includegraphics[width=\textwidth]{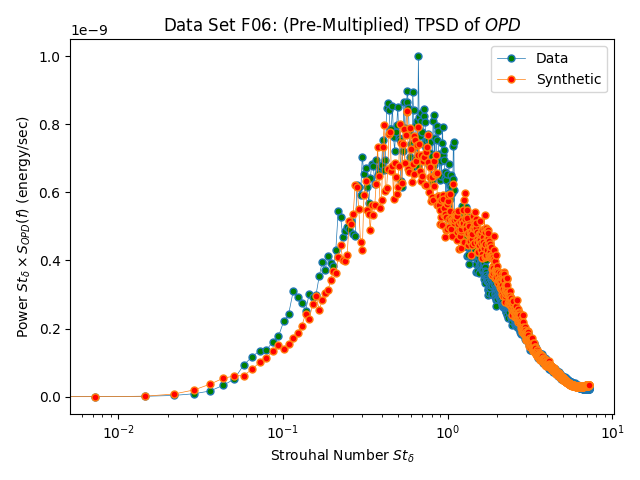}
        \caption{Data Set F06: TPSD of \textit{OPD}}
        \label{fig: F06 TPSD of OPD}
    \end{subfigure}
    \begin{subfigure}{0.45\textwidth}
        \centering
        \includegraphics[width=\textwidth]{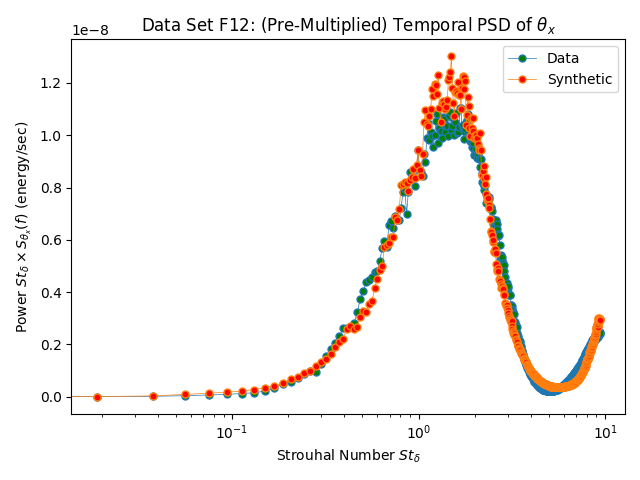}
        \caption{Data Set F12: TPSD of $\theta_x$}
        \label{fig: F12 TPSD of Deflection Angle}
    \end{subfigure}
    \begin{subfigure}{0.45\textwidth}
        \centering
        \includegraphics[width=\textwidth]{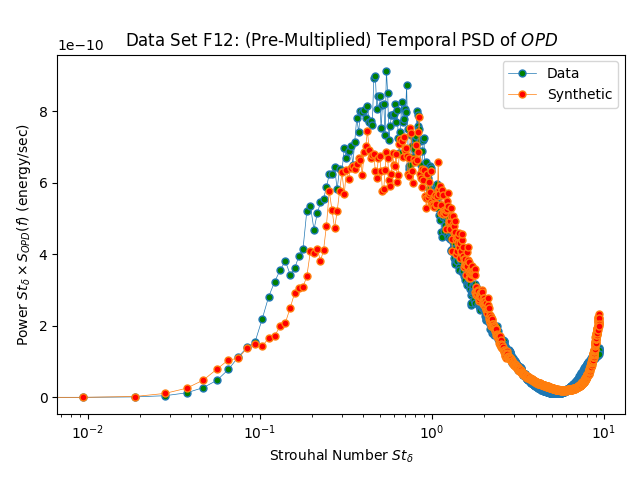}
        \caption{Data Set F12: TPSD of \textit{OPD}}
        \label{fig: F12 TPSD of OPD}
    \end{subfigure}
    \caption{This figure contains TPSD plots calculated from $\theta_x$ (left) and \textit{OPD} (right) for data sets F06 (top) and F12 (bottom). Each graph includes plots of the TPSD calculated from both the original data (blue) and ReVAR output (orange). On the x-axis is Strouhal number $St_{\delta}$ (logarithmic scale) and on the y-axis is pre-multiplied power $St_{\delta} \times S_X(f)$ (linear scale). The quantity $S_X(f)$ is power (energy/sec) as a function of temporal frequency $f$ (Hz). These plots illustrate the close TPSD fit for both $\theta_x$ and \textit{OPD}.}
    \label{fig: TPSD Results}
\end{figure}

In simulating aero-optic effects, we are primarily concerned with accurately capturing the TPSD of the input data. The TPSD of a time-series of data gives the power (energy/sec) as a function of temporal frequency (Hz). The TPSD of measured wavefront aberrations provides an impactful description of aero-optic effects. We thus heavily weigh the TPSD of our synthetic data in determining its description of aero-optics. Further, we were influenced by the use of a SHWFS in \cite{Kemnetz-2016-SanDiego} to calculate TPSD from both \textit{OPD} and the deflection angle $\theta_x=\nabla_x\{OPD\}$ in the stream-wise direction.

Plots of TPSD values show close fit between the measured and synthetic data. The TPSD calculations for both data sets are plotted in Fig.~\ref{fig: TPSD Results} as a function of Strouhal number $St_{\delta}$. We plot the $x$-axis on a logarithmic scale, thus motivating us to pre-multiply the $y$-axis by $St_{\delta}$. This is done for the purpose of visual TPSD integration. Since the $x$-axis is plotted on a logarithmic scale, pre-multiplying the $y$-axis allows us to visually integrate the TPSD from the plot. This method of evaluation shows high-quality results of the ReVAR algorithm since synthetic TPSD has very small error.

\section{CONCLUSIONS}
\label{s: Conclusions}
In this paper, we developed a new aero-optics simulation algorithm: ReVAR, which captures the spatial and temporal statistics of an input time-series of images on arbitrarily long time-scales. We tested ReVAR on TBL data taken from a wind tunnel experiment to assess the algorithm's performance. Our results show that ReVAR captures the TPSD of both  \textit{OPD} and $\theta_x$ in its output synthetic data. Importantly, our implementation of this algorithm is computationally efficient and runs quickly on a standard laptop. The results from this paper verify that ReVAR can efficiently capture the TPSD of aero-optics wavefront aberration data. A future step in this work is testing ReVAR on data from AAOL flight experiments \cite{jumperzenk-grodeyev}, in which laser beams were propagated from a turret on one aircraft and received by a second aircraft. On the second aircraft, a SHWFS measured the resulting aero-optics wavefront aberrations. Because this experiment reflects a realistic aerodynamic environment, AAOL data provides an excellent future testing step for ReVAR.

\appendix    %>>>> this command starts appendixes

\section*{DISCLOSURES}
The views expressed are those of the author and do not necessarily reflect thee official policy or position of the Department of the Air Force, the Department of Defense, or the U.S. Government.  Approved for public release; distribution is unlimited.  Public Affairs release approval \# \PAnumber.

% References
\bibliography{report} % bibliography data in report.bib
\bibliographystyle{spiebib} % makes bibtex use spiebib.bst

\end{document}